\documentclass[prc,twocolumn,floatfix,groupedaddress,nofootinbib,showpacs,preprintnumbers,amsmath,amssymb,amsfonts,superscriptaddress,widetable,href] {revtex4}

\usepackage{footnote}
\usepackage{wasysym}
\usepackage{color}
\usepackage{graphicx}
\usepackage{dcolumn}
\usepackage{multirow}
\usepackage{bm}     
\usepackage{sidecap}
\usepackage{ulem}
\usepackage{hyperref}
\begin{document}

\title{Model dependence of the neutron-skin thickness on the symmetry energy}
 
\author{C. Mondal}
\email{chiranjib.mondal@saha.ac.in}
\address{Saha Institute of Nuclear Physics, 1/AF Bidhannagar, Kolkata {\sl 700064}, India}
\author{B. K. Agrawal}
\email{bijay.agrawal@saha.ac.in}
\address{Saha Institute of Nuclear Physics, 1/AF Bidhannagar, Kolkata {\sl 700064}, India}
\author{M. Centelles}
\address{Departament d'Estructura i Constituents de la Mat\`eria and Institut de Ci\`encies del Cosmos, Facultat de F\'{\i}sica, Universitat de Barcelona, Diagonal 645, E-08028 Barcelona, Spain}
\author{G.  Col\`o}
\address{Dipartimento di Fisica, Universit\`a degli Studi di Milano, via Celoria 16, I-20133 Milano, Italy }
\address{INFN, sezione di Milano, via Celoria 16, I-20133 Milano, Italy}
\author{X. Roca-Maza}
\address{Dipartimento di Fisica, Universit\`a degli Studi di Milano, via Celoria 16, I-20133 Milano, Italy }
\address{INFN, sezione di Milano, via Celoria 16, I-20133 Milano, Italy}
\author{N. Paar}
\address{Department of Physics , Faculty of Science, University of Zagreb, Zagreb, Croatia}
\author{X. Vi\~nas} 
\address{Departament d'Estructura i Constituents de la Mat\`eria and Institut de Ci\`encies del Cosmos, Facultat de F\'{\i}sica, Universitat de Barcelona, Diagonal 645, E-08028 Barcelona, Spain}
\author{S. K. Singh}
\address{Institute of Physics, Bhubhaneshwar, 751005, India}
\author{S. K. Patra}
\address{Institute of Physics, Bhubhaneshwar, 751005, India}


\begin{abstract} 
The model dependence in the correlations of the neutron-skin thickness
in heavy nuclei with various symmetry energy parameters is analyzed by
using several families of systematically varied microscopic mean field
models.  Such correlations show a varying degree of model dependence
once the results for all the different families are combined.  Some mean
field models associated with similar values of the symmetry energy slope
parameter at saturation density $L$, and pertaining to different families,
yield a greater-than-expected spread in the neutron-skin thickness of
the $^{208}$Pb nucleus. The effective value of the symmetry energy slope
parameter $L_{\rm eff}$, determined by using the nucleon density profiles
of the finite nucleus and the density derivative $S^\prime(\rho)$ of the
symmetry energy starting from about saturation density up to low densities
typical of the surface of nuclei, seems to account for the spread in the
neutron-skin thickness for the models with similar $L$. The differences
in the values of $L_{\rm eff}$ are mainly due to the small differences in
the nucleon density distributions of heavy nuclei in the surface region
and the behavior of the symmetry energy at subsaturation densities.

\end{abstract}
\pacs{21.65.Ef, 21.65.Mn, 21.10.Gv} \keywords{Symmetry energy,symmetry energy slope parameter, nuclear matter, neutron skin }

\maketitle

\section{Introduction}

The terrestrial nuclei are mostly asymmetric (i.e., \mbox{$N\not= Z$}),
except for the light nuclei with proton number \mbox{$Z \le 28$}. At
the other extreme, the matter in the compact astrophysical objects like
neutron stars is highly asymmetric \cite{Shapiro83}. The asymmetry
in the finite nuclei primarily arises due to the balance between
the Coulomb energy and the nuclear symmetry energy. The conditions of
$\beta-$equilibrium and charge neutrality render the matter in a neutron
star to be highly asymmetric or predominantly composed of neutrons
\cite{Glendenning00}. The densities at the center of nuclei are close to
the normal saturation density $\rho_0$ ($ 0.16\ \text{fm}^{-3}$), whereas
the densities at the center of neutron stars are predicted to be typically
a few times $\rho_0$. Thus, the accurate knowledge of the nuclear symmetry
energy over a wide range of densities is indispensable to understand a
variety of phenomena in finite nuclei as well as in neutron stars.

The details of the density dependence of the nuclear symmetry
energy remain hard to isolate, though progress in this
direction has been made in the last few years (see for instance
Refs.\cite{Furnstahl02, Danielewicz03, Steiner05, Centelles09, Warda09,
Piekarewicz12,Agrawal12,Roca-Maza13, Roca-Maza13a, Roca-Maza11, Tsang12,
Centelles10, EPJA14, Chen05, Mondal15, Reinhard16, Roca-Maza15, Li08}
and the experimental and theoretical works quoted therein). The density
dependence of the nuclear symmetry energy around saturation is governed
to leading order by its density derivative expressed as
\begin{eqnarray}
L=3\rho_0\left (\frac{d S(\rho)}{d\rho}\right)_{\rho_0},
\label{eq:L}
\end{eqnarray}
where $S(\rho)$ is the symmetry energy at a density $\rho$.
The macroscopic nuclear droplet model (DM) of Myers and Swiatecki
\cite{Myers69, Myers80} suggests that various symmetry energy parameters
and the neutron-skin thickness in a heavy nucleus are related to one
another. The neutron skin thickness is defined as the difference between
the rms radii for the density distributions of the neutrons and protons
in the nucleus:
\begin{equation}
\Delta r_{\rm np} \equiv \langle r^2 \rangle_n^{1/2} 
- \langle r^2 \rangle_p^{1/2}.
\label{skin}
\end{equation} 
Nuclear mean-field models predict a nearly linear correlation of $\Delta
r_{\rm np}$ of a heavy nucleus such as $^{208}$Pb with the slope of the
equation of state of neutron matter at a subsaturation density around
0.1 fm$^{-3}$ \cite{Brown00a,Brown01}, with the density derivative
of the symmetry energy $L$ \cite{Furnstahl02,Centelles09,Warda09,
Danielewicz03,Avancini07,Vidana09, Chen05}, and with the surface symmetry
energy in a finite nucleus \cite{Danielewicz03,Centelles09,Satula06}.
The correlation of a finite nucleus property such as $\Delta r_{\rm
np}$ with a bulk property of infinite nuclear matter such as $L$ can be
interpreted as basically due to the dependence of $\Delta r_{\rm np}$
on the surface symmetry energy.  In a local density approximation
the surface symmetry energy can be correlated with $L$, and this
fact therefore implies the correlation between $\Delta r_{\rm np}$
and $L$. Macroscopic approaches such as the DM \cite{Myers69,Myers80}
often provide insightful guidance into the global features of many of
these correlations \cite{Centelles09,Warda09,Centelles10}, as it will
be briefly recalled in the next section.

The Lead Radius Experiment (PREX) \cite{Horowitz01, Abrahamya12}
has recently measured the neutron skin thickness $\Delta r_{\rm np}$
of $^{208}$Pb.  This experiment is performed via parity-violating
electron scattering \cite{Donnelly89}  and provides the first
purely electroweak, model independent measurement of the weak charge
form factor, closely connected to the neutron distribution of the
$^{208}$Pb nucleus \cite{Donnelly89}. By measuring the weak form
factor of $^{208}$Pb at momentum transfer $q\approx 0.475$ fm$^{-1}$,
PREX was able to determine $\Delta r_{\rm np} = 0.33^{+0.16} _{-0.18}$
fm \cite{Abrahamya12}.  Recently, a follow-up measurement of PREX has
been proposed which intends to measure the neutron-skin thickness in
the $^{208}$Pb nucleus with an accuracy of $0.06$ fm  \cite{Jlab}.
The hadronic probes are also used to estimate the neutron distribution
in nuclei \cite{Hoffmann80,Zenihiro10,Krasznahorkay04,Klos07,Friedman09}.
In this case, the strong interaction needs to be modeled and, therefore,
deducing the neutron radius from these experiments can imply various
theoretical uncertainties, which in some cases are difficult to estimate.
The analyses from recent hadronic experiments have led to varying
values of the neutron skin thickness of $^{208}$Pb, $\Delta r_{\rm np} =
0.16\pm 0.02$(stat)$\pm0.04$(syst) fm \cite{Klos07}  and $\Delta r_{\rm
np} = 0.211^{+0.054}_{-0.063}$ fm \cite{Zenihiro10}. A very recent
measurement of coherent pion photo-production \cite{Tarbert14} provides
a value $\Delta r_{\rm np} = 0.15 \pm 0.03$ fm  for $^{208}$Pb.  Also a
neutron skin thickness $\Delta r_{\rm np} = 0.165\pm 0.009$(exp)$\pm
0.013$(theor)$\pm 0.021$(est) fm has been extracted recently from
comparison of theory with the measured electric dipole polarizability in
$^{208}$Pb \cite{Roca-Maza13, Roca-Maza15, Tamii11, Hashimoto15, Rossi13}.

Ongoing efforts are underway to perform an accurate and model independent
measurement of the neutron-skin thickness in the $^{208}$Pb nucleus. At
the same time, it may not be straightforward for theory to extract
various symmetry energy parameters from the neutron-skin thickness in a
model-independent fashion. Starting from the seminal papers of more than
a decade ago \cite{Brown00a,Brown00,Brown01,Furnstahl02}, the focus has
mainly been on the linear correlation between the neutron-skin thickness
and the slope parameter $L$ of the symmetry energy. The correlation
is satisfied to a large degree in the microscopic calculations
with mean field models but it is not perfect and a certain model
dependence appears in the results (see for example the plots in Refs.
\cite{Brown00,Brown01,Furnstahl02,Danielewicz03,Centelles09,Warda09,
Centelles10}).  By model dependence we mean here that different mean-field
models may predict similar values for the $L$ parameter but different
neutron skin thickness in a heavy nucleus. As it may be seen for example
from Fig. 2 and Table II of Ref. \cite{Centelles10}, some models deviate
from the linear correlation. This analysis was done by using different
unbiasedly selected mean-field models. We would like to complement the
earlier analysis with the one based on families of systematically varied
models, in an attempt to identify the sources for the model dependence
in the correlations.

In the present work we revisit the correlations of $\Delta r_{\rm
np}$ with various symmetry energy parameters. The plausible causes
for the existence of a model dependence in these correlations are
investigated. The correlations are evaluated by using five different
families of systematically varied microscopic mean-field models. Three
out of these five families correspond to relativistic energy density
functionals \cite{Serot86,Boguta77}  and the remaining two families
correspond to a non-relativistic functional \cite{Vautherin72}. We also
predict the neutron skin thickness of the neutron-rich nucleus $^{132}$Sn
which has not been measured yet.

The paper is organized as follows. The geometrical definitions employed to
decompose the neutron-skin thickness into bulk and surface contributions
\cite{Warda10,Centelles10} are briefly outlined in Sec. II. We also
provide in  this section some results derived from the macroscopic
DM suggesting possible connections between the neutron-skin thickness
and various symmetry energy parameters. In Sec. III, the results for
the correlations of the neutron-skin thickness in the $^{208}$Pb and
$^{132}$Sn nuclei with the symmetry energy parameters obtained for several
families of the systematically varied models are presented. The plausible
causes for the model dependence in such correlations are investigated
in detail. The main conclusions are presented in Sec. IV.

\section{Neutron-skin thickness and symmetry energy parameters}
\label{theo}

From a geometrical point of view, the neutron skin thickness in a nucleus
may be thought as originated by two different effects. One effect is
due to the separation between the mean sharp surfaces of the neutron
and proton density distributions. Since this effect corresponds to a
different extent of the bulk region of the neutron and proton densities,
we refer to it as the bulk contribution to the neutron skin thickness. The
other effect is due to the different surface widths of the neutron and
proton densities, which we call the surface contribution to the neutron
skin thickness. To compute the bulk and surface contributions to the
neutron skin thickness in a nucleus requires a proper definition of
these quantities based on the nuclear densities. In this respect we
follow closely the method described by Hasse and Myers \cite{Hasse88}
and which we applied in Refs. \cite{Warda10,Centelles10}.

In order to determine the position of the neutron and proton effective
surfaces one can define different radii. In particular, one can define
the central radius $C$ as
\begin{equation}
\label{c}
C= \frac{1}{\rho(0)} 
\int_0^{\infty} \rho(r) dr .
\end{equation}
Another option for the mean position of the surface is the equivalent
radius $R$, which is the radius of a uniform sharp distribution whose
density equals the bulk value of the actual density and has the same
number of particles:
\begin{equation}
\label{r}
\frac 43 \pi R^3 \rho({\rm bulk}) =
4\pi \int_0^{\infty} \rho(r) r^2 dr \,.
\end{equation}
Finally, one can also define the  equivalent rms radius $Q$ that
describes a uniform sharp distribution with the same rms radius as the
given density:
\begin{equation}
\label{q}
\frac 35 \, Q^2= \langle r^2 \rangle \,.
\end{equation}

The radii $C$, $R$, and $Q$ are related by the expressions \cite{Hasse88}
\begin{equation}
\label{qr}
Q= R\left(1+\frac{5}{2}\frac {b^2}{R^2}+ ...\right) \quad
C= R\left(1-\frac {b^2}{R^2}+ ...\right),
\end{equation}
where $b$ is the surface width of the density profile defined as
\begin{equation}
\label{b}
b^2= - \frac{1}{\rho(0)}
\int_0^{\infty} (r-C)^2 \frac{d \rho(r)}{dr} dr,
\end{equation}
which provides a measure of the extent of the surface of the nucleus.
The neutron skin thickness, which is defined through the rms radii, can 
be expressed by
\begin{equation}
\label{r0}
\Delta r_{np}=\sqrt{\frac{3}{5}} \left(Q_n-Q_p\right) ,
\end{equation}
and using Eq.(\ref{qr}) reads:
\begin{equation}
\Delta r_{\rm np} = \sqrt{\frac{3}{5}} \left[\left(R_n - R_p\right)
+ \frac{5}{2}\left(\frac{b_n^2}{R_n}-\frac{b_p^2}{R_p}\right)\right],
\label{rnptotal}
\end{equation}
which clearly separates the bulk and surface contributions as
 \begin{equation}
\Delta r_{\rm np}^{\rm bulk} \equiv \sqrt{\frac{3}{5}} \left(R_n - R_p\right),
\label{rnpbulk}
\end{equation}
and 
\begin{equation}
\Delta r_{\rm np}^{\rm surf} \equiv\sqrt{\frac{3}{5}} \frac{5}{2}
\left(\frac{b_n^2}{R_n}-\frac{b_p^2}{R_p}\right).
\label{rnpsurf}
\end{equation}
In Eqs. (\ref{rnptotal}) and (\ref{rnpsurf}), we have neglected
$\mathcal{O}\left[b^4/R^3\right]$ and higher-order terms since they
represent a small correction \cite{Centelles10} to $\Delta r_{\rm
np}$---of less or around a 1-2\%---that will leave our conclusions
unchanged.

In order to extract the bulk and surface contributions to the neutron
skin thickness from the quantal proton and neutron densities obtained
within the Skyrme Hartree-Fock or the relativistic mean-field models,
we proceed as in Refs.\cite{Centelles10, Warda10}. That is, we fit the
self-consistent quantal proton and neutron densities by two-parameter
Fermi (2pF) distributions
\begin{equation}
\rho_q(r)=\frac{\rho_{0,q}}{1+{\rm exp}[(r-C_q)/a_q]},
\label{2pF}
\end{equation}
where $q=n,p$. The parameters $\rho_{0,q}$,  $C_q$ and $a_q$ are adjusted
to reproduce the nucleon numbers as well as the values for the second
and fourth moments of the actual density distributions, i.e.,  $\langle
r^2_q\rangle$ and $\langle r^4_q\rangle$.  Once this fit is done,
we can express Eqs. (\ref{rnptotal})--(\ref{rnpsurf}) for the neutron
skin thickness in terms of the parameters $C_q$ and $a_q$ taking into
account Eq.(\ref{qr}) and the fact that for a 2pF distribution $b=\pi
a/\sqrt{3}$. Therefore, the bulk and surface contributions to the neutron
skin thickness can be written as
\begin{equation}
\Delta r_{\rm np}^{\rm bulk}= \sqrt{\frac{3}{5}}
\left[(C_n-C_p) +\frac{\pi^2}{3}\left(\frac{a_n^2}{C_n}-
\frac{a_p^2}{C_p}\right)\right ], 
\label{blk}
\end{equation}
\begin{equation}
\Delta r_{\rm np}^{\rm surf}=  \sqrt{\frac{3}{5}}\frac {5 \pi^2}{6}
\left(\frac{a_n^2}{C_n}-\frac{a_p^2}{C_p}\right),
\label{srf}
\end{equation}
up to terms of order $\mathcal{O}\left[a^4/C^3\right]$. It should
be mentioned that, the $\Delta r_{\rm np}$ values calculated from the
actual densities obtained self consistently match very well with the ones
calculated by summing Eqs. (\ref{blk}) and (\ref{srf}) after applying
our prescription to determine the parameters of the Fermi function.

Some insight about possible correlations between the neutron skin
thickness and different observables related to the symmetry energy is
provided by the DM \cite{Myers80}. Within this model, which neglects
shell correction effects, the neutron skin thickness is expressed by
\begin{equation}
\Delta r_{\rm np} = \sqrt{\frac{3}{5}} \left[t - \frac{e^2Z}{70 J}
+ \frac{5}{2R}\left(b_n^2 - b_p^2\right)\right],
\label{rnpdm}
\end{equation}
where $e^2Z/70J$ is a correction due to the Coulomb interaction, $R=r_0
A^{1/3}$ is the nuclear radius, and $b_n$ and $b_p$ are the surface
widths of the neutron and proton density profiles.  The quantity $t$
in (\ref{rnpdm}) represents the distance between the location of the
neutron and proton mean surfaces and therefore is proportional to the
bulk contribution to the neutron skin thickness. In the DM its value is
given by
\begin{equation}
t = \frac{3}{2}r_0 \frac{J}{Q_{\rm stiff}} \frac{I-I_C}{1 + x_A}, 
\label{t}
\end{equation}
with 
\begin{equation}
I_C = \frac{3e^2}{5r_0}\frac{Z}{12J}A^{-1/3} \quad  \text{and} \quad 
x_A = \frac{9J}{4Q_{\rm stiff}}A^{-1/3},
\label{ICXA}
\end{equation}
where $I=(N-Z)/A$, $J$ is the bulk symmetry energy at saturation, and
$Q_{\rm stiff}$ is the surface stiffness. For each mean field model,
the parameters $r_0$ and $J$ can be obtained from calculations in
infinite nuclear matter and $Q_{\rm stiff}$ from calculations performed
in semi-infinite nuclear matter \cite{Warda09,Centelles98,Estal99}.

Within the DM, the symmetry energy coefficient of a finite nucleus of
mass number $A$ is given by
\begin{equation}
a_{sym}(A) = \frac{J}{1 + x_A}.
\label{symdm}
\end{equation}
Replacing $a_{sym}(A)$ in Eq. (\ref{t}), the separation distance between
the mean surfaces of neutrons and protons can be recast as
\begin{equation}
t = \frac{2r_0}{3J}[J - a_{sym}(A)]A^{1/3} (I -I_C). 
\label{t1}
\end{equation}
The link between a property in finite nuclei such as $a_{sym}(A)$ and some
symmetry energy parameters in infinite nuclear matter may be obtained from
the observation \cite{Centelles09} that for a heavy nucleus there is a
subsaturation density, which for $^{208}$Pb is around 0.1 fm$^{-3}$, such
that the symmetry energy coefficient in the finite nucleus $a_{sym}(A)$
equals the symmetry energy in nuclear matter $S(\rho)$ computed at that
density. This relation is roughly independent of the mean field model
used to compute it. Around the saturation density $\rho_0$ the symmetry
energy can be expanded as
\begin{equation}
S(\rho) \simeq J -  L\Big(\frac{\rho_0-\rho}{3\rho_0}\Big) 
+ \frac{1}{2}K_{sym}\Big(\frac{\rho_0-\rho}{3\rho_0}\Big)^2.
\label{s_rho}
\end{equation}
Consequently, the distance $t$ can be finally expressed approximately as
\cite{Centelles09}
\begin{equation}
t = \frac{2r_0}{3J}L\Big(\frac{\rho-\rho_0}{3\rho_0}\Big)
\Big[1 - \frac{K_{sym}}{2L} \Big(\frac{\rho-\rho_0}{3\rho_0}\Big) \Big] A^{1/3} (I -I_C).
\label{t2}
\end{equation}
Equations (\ref{t1}) and (\ref{t2}) suggest correlations between
the bulk neutron skin thickness in finite nuclei and some isovector
indicators such as $J-a_{sym}(A)$, $a_{sym}(A)/J$ and $L$, which will be
discussed in detail along this paper. To compute the average symmetry
energy of a finite nucleus with the DM (Eq. (\ref{symdm})) requires
the knowledge of the surface stiffness $Q_{\rm stiff}$, which in turn
requires semi-infinite nuclear matter calculations \cite{Warda09}. An
efficient procedure to circumvent this, is to evaluate $a_{sym}(A)$
within a local density approximation as \cite{Agrawal12}
\begin{eqnarray}
a_{\rm sym}(A)=\frac{4 \pi}{AI^2}\int \ [r^2 \rho(r)I^2(r)] S(\rho(r))dr,
\label{eq:Asym}
\end{eqnarray}
where $I(r) = \frac{\rho_n(r) -\rho_p(r)}{\rho(r)}$ is the local isospin
asymmetry and $\rho (r)$ is the sum of the neutron and proton densities.
This approximation works very well for medium heavy $^{132}$Sn or heavy
$^{208}$Pb nuclei \cite{Liu13}.

\section{Results and discussions}
The neutron-skin thickness and several symmetry energy parameters are
calculated using five different families of systematically varied
models, namely, the \mbox{SAMi-J} \cite{Roca-Maza12, Roca-Maza13},
DDME \cite{Vretenar03}, FSV, TSV and KDE0-J models. The energy density
functional associated with DDME, FSV, and TSV corresponds to an effective
Lagrangian density typical of the relativistic mean-field models, whereas
SAMi-J and KDE0-J are based on the standard form of the Skyrme force.

We have obtained the different families of systematically varied
parameter sets so that they explore different values of the symmetry
energy parameters around an optimal value, while reasonably keeping the
quality of the best fit.  The values of the neutron-skin thickness in a
heavy nucleus like $^{208}$Pb vary over a wide range within the families
due to the variations of the symmetry energy parameters. The parameter
sets for the FSV, TSV and KDE0-J families are obtained in the present
work. The effective Lagrangian density employed for the FSV family is
similar to that for the FSU model \cite{Todd-Rutel05}.  In addition to
the coupling of $\rho$ meson to the nucleons as conventionally employed,
the presence of a cross-coupling between the $\omega$ and $\rho$ mesons in
the FSU model enables one to vary the symmetry energy, and accordingly
the symmetry energy slope parameter $L$, over a wide range without
significantly affecting the quality of the fit to the bulk properties
of the finite nuclei. The TSV family is obtained using the effective
Lagrangian density as introduced in Ref. \cite{Estal01} in which the
$\rho-$meson and its coupling to the $\sigma-$meson govern the isovector
part of the interactions between the nucleons. The $\omega-\rho$ cross
coupling in the FSV family and the $\sigma-\rho$ cross coupling in the
TSV family produce different behaviors in the density dependence of
the symmetry energy, because the source term for the $\omega$-field
is governed by the baryon density and that for the $\sigma$-field
is governed by the scalar density. The experimental data employed to
determine the TSV and FSV families are the total binding energies for
the $^{16}{\rm O}, ^{40,48}{\rm Ca}, ^{68}{\rm Ni}, ^{90}{\rm Zr},
^{100,132}{\rm Sn}, ^{208}{\rm Pb}$ nuclei, and the root mean square
charge radii for the $^{16}{\rm O}, ^{40,48}{\rm Ca}, ^{90}{\rm Zr},
^{208}{\rm Pb}$ nuclei. The energy density functional for the KDE0-J
family calculated within the Skyrme ansatz is taken from the KDE0 force
of Ref. \cite{Agrawal05}.  The model parameters are constrained to
yield the nuclear matter incompressibility coefficient in the range of
225--250 MeV. The calculated values of the total binding energy and the
charge radius for the $^{208}$Pb nucleus obtained for all the models
considered deviate from the experimental data only within $0.25\%$
and $0.8\%$, respectively.

\begin{figure}
\includegraphics[height=3.4in,width=3.2in,angle=-90]{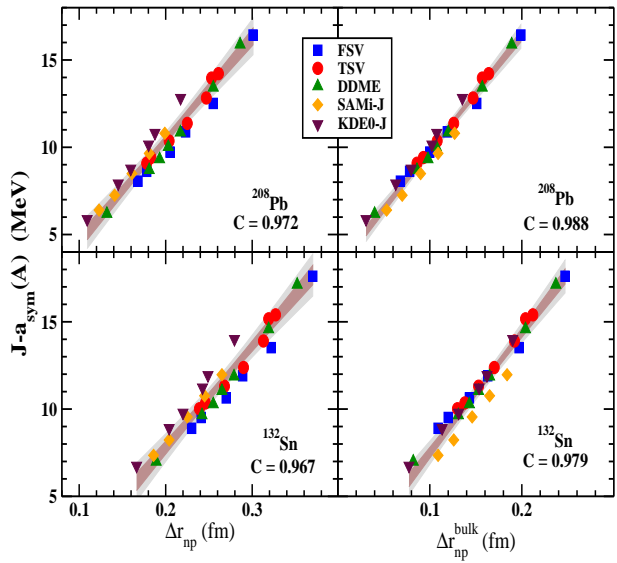}
\caption{\label{fig1} (Color online) Plots for the difference between the
symmetry energy coefficient for  infinite nuclear matter $J$ and that
for finite nuclei $a_{\rm sym}(A)$  as a function of the neutron-skin
thickness (left panels) and of the bulk part of the neutron-skin thickness
(right panels).  The results are obtained using five different families
of mean-field models, namely, FSV (blue squares), TSV (red circles), DDME
(green triangles), SAMi-J (orange diamonds) and KDE0-J (maroon inverted
triangles). The correlation coefficients are: $C(J - a_{\rm sym}(A),
\Delta r_{\rm np} )$ = 0.972 (0.967) and  $C(J- a_{\rm sym}(A), \Delta
r_{\rm np}^{\rm bulk})$ = 0.988 (0.979)  for  $^{208}$Pb ($^{132}$Sn)
nuclei.  The inner (outer) colored regions depict the loci of the 95\%
confidence (prediction) bands of the regression (see, e.g., Chap. 3
of Ref.\cite{Draper81}).}
\end{figure}

\subsection{Correlation plots associated with isovector indicators}

\begin{figure}{}
\includegraphics[height=3.5in,width=3.2in,angle=-90]{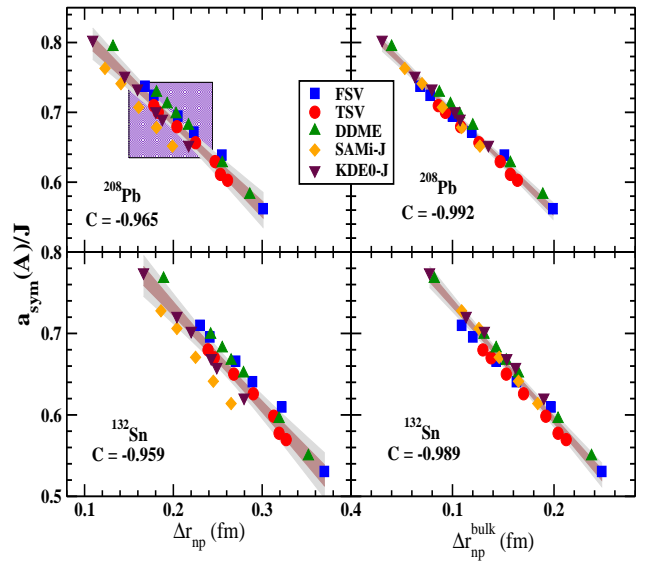}
\caption{\label{fig2} (Color online) Plots for the ratio of the  nuclear
symmetry energy coefficient for finite nuclei $a_{\rm sym}(A)$ to that for
infinite nuclear matter $J$, as a function of the neutron-skin thickness
(left panels) and of the bulk part of the neutron-skin thickness (right
panels).  The square shaded region in the upper-left panel corresponds to
$ a_{\rm sym}(A) = 22.4\pm 0.3$ MeV and $J= 32.5 \pm 2.5$ MeV. 
The correlation coefficients are $|C(a_{\rm sym}(A)/J,
\Delta r_{\rm np} )|$ = 0.965 (0.959) and  $|C(a_{\rm sym}(A)/J, \Delta
r_{\rm np}^{\rm bulk} )|$ = 0.992 (0.989)  for $^{208}$Pb ($^{132}$Sn)
nuclei.  The inner (outer) colored regions depict the loci of the 95\%
confidence (prediction) bands of the regression (see, e.g., Chap. 3
of Ref.\cite{Draper81}).}
\end{figure}

As we discussed in the previous Section, the DM is a useful guideline to
suggest the kind of correlations that we can expect between the neutron
skin thickness and the symmetry energy parameters. As shown in Ref.
\cite{Centelles10}, these correlations are mainly due to the bulk term
of Eq.(\ref{rnpdm}) rather than to the surface contribution to $\Delta
r_{\rm np}$. In the bulk part of $\Delta r_{\rm np}$, the quantity $\left(
J-a_{\rm sym}(A) \right)/J$ determines the ratio of the surface symmetry
to volume symmetry energies, see Eq.(\ref{t1}); the close relation
of different isovector observables in finite nuclei with the ratio of
the surface and volume symmetry energies has been observed in several
studies, cf. for example Refs. \cite{Satula06,Roca-Maza13a} and references
therein. The values of $r_0$ for the various models considered in the
present work display only a small variation indicating that the total
neutron-skin thickness $\Delta r_{\rm np}$ of a given heavy nucleus may
be correlated to the ratio $\left( J-a_{\rm sym}(A) \right)/J$, or also
to the difference $(J-a_{\rm sym}(A))$ provided the value of $J$ does not
show a large variation as compared to $\left( J-a_{\rm sym}(A) \right)$.

In Fig. \ref{fig1}, we plot for the $^{208}$Pb and $^{132}$Sn nuclei
the values of $J - a_{\rm sym}(A)$ as a function of $\Delta r_{\rm np}$
in the left panel, and as a function of the bulk part of the neutron-skin
thickness $\Delta r_{\rm np}^{\rm bulk}$ in the right panel. The results
are reported for the five different families of systematically varied
models, namely, FSV, TSV, SAMi-J, DDME and KDE0-J as indicated in the
figure. Fairly evident linear correlations are observed between $J -
a_{\rm sym}(A)$ and both $\Delta r_{\rm np}$ and $\Delta r_{\rm np}^{\rm
bulk}$. More quantitatively, if we calculate the Pearson's correlation
coefficients $C(X,Y)$ \cite{Brandt97}, their values are $C(J-a_{\rm
sym}(A),\Delta r_{\rm np})$ = 0.972 (0.967) and $C(J-a_{\rm sym}(A),
\Delta r_{\rm np}^{\rm bulk})$ = 0.988 (0.979) for the $^{208}$Pb
($^{132}$Sn) nuclei, respectively.  Thus, the correlation of $J-a_{\rm
sym}(A)$ with $\Delta r_{\rm np}^{\rm bulk}$ is a little higher than
with $\Delta r_{\rm np}$ for both $^{208}$Pb and $^{132}$Sn nuclei,
as it may be expected from the discussions in Sec. \ref{theo}.

Following Eq. (\ref{t1}) one can directly correlate $\left( J-a_{\rm
sym}(A) \right)/J$ (or equivalently $a_{\rm sym}(A)/J$) with $\Delta
r_{\rm np}$ of a heavy nucleus.  In Fig. \ref{fig2} we display the
ratio $a_{\rm sym}(A)/J$ as a function of $\Delta r_{\rm np}$ and of
$\Delta r_{np}^{\rm bulk}$ for the $^{208}$Pb and $^{132}$Sn nuclei. The
correlations of $a_{\rm sym}(A)/J$ with $\Delta r_{\rm np}$ are relatively
weaker in comparison to those with $\Delta r_{\rm np}^{\rm bulk}$. In
the case of $a_{\rm sym}(A)/J$ and $\Delta r_{\rm np}$ the correlation
coefficient is $|C(a_{\rm sym}(A)/J, \Delta r_{\rm np} )| = 0.965$ (0.959)
for $^{208}$Pb ($^{132}$Sn), whereas in the case of $a_{\rm sym}(A)/J$
and $\Delta r_{\rm np}^{\rm bulk}$ the correlation coefficient increases
up to high values $|C(a_{\rm sym}(A)/J, \Delta r_{\rm np}^{\rm bulk} )|
= 0.992$ (0.989) for $^{208}$Pb ($^{132}$Sn).

At this point, it is interesting to address the constraints on the
neutron-skin thickness that may be deduced from the present study. The
rectangular shaded region in the upper-left panel of Fig. \ref{fig2}
corresponds to $ a_{\rm sym}(A) = 22.4\pm 0.3$ MeV for $^{208}$Pb
\cite{Fan14} and $J= 32.5 \pm 2.5$ MeV, which yields $\Delta r_{\rm
np}= 0.197 \pm 0.047$ fm in the $^{208}$Pb nucleus.  The constraint
$a_{\rm sym}(A)= 22.4\pm 0.3$ MeV was evaluated in Ref. \cite{Fan14}
using the experimental binding energy differences.  Furthermore, the
effect of the Coulomb interaction on the surface asymmetry and the
effect of the surface diffuseness on the Coulomb energy were taken into
account. The value of $J = 32.5\pm 2.5$ MeV as used in the present work
has a quite reasonable overlap with the ones extracted either from a
version of the finite-range droplet model (FRDM) that performs very
well in reproducing the experimental mass systematics \cite{Moller12},
by analyzing the experimental data on the electric dipole polarizability
in  $^{68}$Ni, $^{120}$Sn and $^{208}$Pb nuclei \cite{Roca-Maza15}, from
specific manipulation of the semi-empirical mass formula \cite{Jiang12},
through analysis of the properties of semi-infinite nuclear matter
\cite{Danielewicz09}, or by analyzing pygmy dipole resonance data on
$^{68}$Ni and $^{132}$Sn nuclei \cite{Carbone10}. This value of $J$ also
overlaps with the conclusions provided in recent papers \cite{Tsang12,
Lattimer13}.

\begin{figure}
\includegraphics[height=3.5in,width=3.2in,angle=-90]{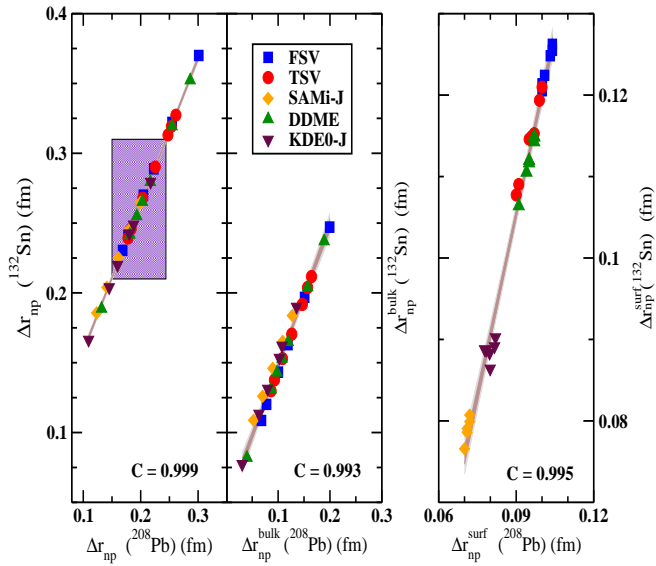}
\caption{\label{fig3} (Color online) 
Neutron-skin thickness (left) and its bulk (middle) and surface
(right) contributions for the $^{132}$Sn nucleus plotted against
the same quantities for the $^{208}$Pb nucleus.  The shaded region
corresponds to the values of the neutron-skin thickness in $^{132}$Sn
determined from the ones estimated for the $^{208}$Pb nucleus (see also
Fig. \ref{fig2}).  The correlation coefficients obtained for the results
presented in the left, middle  and right panels are 0.999, 0.993 and
0.995, respectively.  The inner (outer) colored regions depict the loci
of the 95\% confidence (prediction) bands of the regression (see, e.g.,
Chap. 3 of Ref.\cite{Draper81}).}
\end{figure}

It is desirable to check the degree of consistency between the results
for different heavy nuclei, in particular between $^{208}$Pb and
$^{132}$Sn which would allow to predict the neutron skin thickness of
the nucleus $^{132}$Sn assumed that the one of $^{208}$Pb is known. In
the left panel of Fig. \ref{fig3}, we plot $\Delta r_{\rm np}$ for the
$^{132}$Sn nucleus against that for the $^{208}$Pb nucleus. Similarly,
the results for $\Delta r_{\rm np}^{\rm bulk}$ and $\Delta r_{\rm np}^{\rm
surf}$ are plotted in the middle and right panels of Fig. \ref{fig3},
respectively. It is observed that the values of $\Delta r_{\rm np}$,
$\Delta r_{\rm np}^{\rm bulk}$ and $\Delta r_{\rm np}^{\rm surf}$ for
the $^{132}$Sn nucleus are very well correlated with the corresponding
values in the $^{208}$Pb nucleus. This is in harmony with earlier work
\cite{Piekarewicz12}. Hence, the information provided by the neutron skin
of two heavy nuclei on the isovector channel of the nuclear effective
interaction is mutually inclusive. Such an observation allows us to
predict $\Delta r_{\rm np}= 0.260 \pm 0.050\,{\rm fm}$ for $^{132}$Sn
nucleus by using the above estimated value for ${}^{208}$Pb of $\Delta
r_{\rm np}= 0.197 \pm 0.047$ fm.

\begin{figure}
\includegraphics[height=3.5in,width=3.2in,angle=-90]{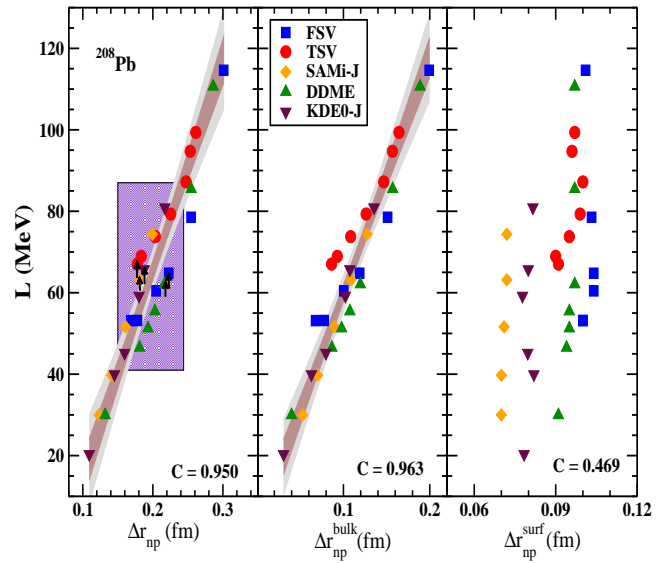}
\caption{\label{fig4} (Color online) Plots for the symmetry energy slope
parameter $L$ as a function of the neutron-skin thickness (left), its bulk
part (middle) and its surface part (right) for the $^{208}$Pb nucleus.
The shaded region in the left panel projects out the values of $L=
64 \pm 23$ MeV obtained from $\Delta r_{\rm np} = 0.197 \pm 0.047$
fm which, in turn, is obtained  by using the empirical values of $J$
and $a_{\rm sym}(A)$ (see also Fig. \ref{fig2}).  The arrow marks in the
left panel indicate the points with the slope parameter $L \sim$ 65 MeV.
 The  values of the correlation coefficients are
$C(L,\Delta r_{\rm np})$ = 0.950, $C(L,\Delta r_{\rm np}^{\rm bulk})$ =
0.963 and $C(L,\Delta r_{\rm np}^{\rm surf})$ = 0.469.  The inner (outer)
colored regions depict the loci of the 95\% confidence (prediction)
bands of the regression (see, e.g., Chap. 3 of Ref.\cite{Draper81}).}
\end{figure}

As recalled above, and discussed in the literature (cf., in particular,
Ref. \cite{Centelles10} and references therein), we expect that
the correlation between the neutron-skin thickness and $\left(
J - a_{\rm sym}(A) \right) / J$ leads to a correlation between the
neutron-skin thickness and the symmetry energy slope parameter $L$. In
Fig. \ref{fig4}, we display the variation of $L$ as a function of $\Delta
r_{\rm np}$ (left), $\Delta r_{\rm np}^{\rm bulk}$ (middle) and $\Delta
r_{\rm np}^{\rm surf}$ (right panel) for the $^{208}$Pb nucleus in the
analyzed families of models.  Using the constraint on $\Delta r_{\rm np}$
($^{208}$Pb) obtained in Fig.  \ref{fig2}, the bound on the value of $L$
comes out to be $L=\ 64\pm23$ MeV; displayed as the shaded region of
left panel in Fig. \ref{fig4}.  The correlation coefficients of $L$ with
$\Delta r_{\rm np}$ and with $\Delta r_{\rm np}^{\rm bulk}$ are lower
than in the case of the correlations displayed in Figs. \ref{fig1} and
\ref{fig2}, suggesting that the neutron-skin thickness is slightly better
correlated with $J - a_{\rm sym}(A)$ or the ratio $a_{\rm sym}(A)/J$ than
with the slope parameter $L$.  This might be a feature of the families we
have chosen and does not necessarily apply to the situation in which one
employs a large set of unbiasedly selected models \cite{Centelles10}. As
above, the $\Delta r_{\rm np}$ - $L$ correlation is weaker in comparison
to the $\Delta r_{\rm np}^{\rm bulk}$ - $L$ correlation, in qualitative
agreement with Ref. \cite{Centelles10}.

The ``arrow''marks in Fig. \ref{fig4} indicate the five models, each
from a different family, with $L$ varying in a narrow range of 62.1
MeV to 67.0 MeV. For these five models, there happens to be a spread in
$\Delta r_{\rm np}$ of almost $ 0.05$ fm which is larger than expected.
In comparison, the equation of the linear fit of the results of all
models in the left panel of Fig. \ref{fig4} gives a variation in the
value of $\Delta r_{\rm np}$ ($^{208}$Pb) with the change of $L$ as,
$\delta (\Delta r_{\rm np})\simeq 0.002\,\delta L$, so that a change in
$L$ of 5 MeV implies an average change in $\Delta r_{\rm np}$ of about
0.01 fm only, which is smaller than the observed spread of 0.05 fm in
the five models mentioned above. The DM supports a similar conclusion,
as it can be seen from Eq. (\ref{t2}) that the DM predicts an average
variation of $\Delta r_{\rm np}$ ($^{208}$Pb) with $L$ approximately as,
$\delta (\Delta r_{\rm np})\simeq 0.003\,\delta L$.  The two mentioned
models from the TSV and SAMi-J families have $L=67$ MeV and $L=63.2$ MeV,
respectively, and yield in $^{208}$Pb smaller values of $\Delta r_{\rm np}
\simeq 0.18$ fm, whereas the two models from the FSV and DDME families
have $L=64.8$ MeV and $L=62.1$ MeV, respectively, and give rise to larger
values of $\Delta r_{\rm np} \simeq 0.22$ fm. The model from KDE0-J family
with $L=65.7$ MeV yields an intermediate value of $\Delta r_{\rm np}$
($^{208}$Pb) $\simeq 0.19$ fm.  Actually, it comes as an intriguing fact
that the extracted values of $\Delta r_{\rm np}$ differ by $\sim0.05$
fm for the two models of the FSV and TSV families with similar $L$,
although the parameters for these two families are obtained by using
exactly the same kind of fitting protocol. In the next subsection, we
aim to search for plausible interpretations for such differences in the
neutron skin thickness corresponding to models with similar $L$ values.

\subsection{Systematic differences between the families of functionals}
\begin{figure}[t]
{\includegraphics[height=3.5in,width=3.2in,angle=-90]{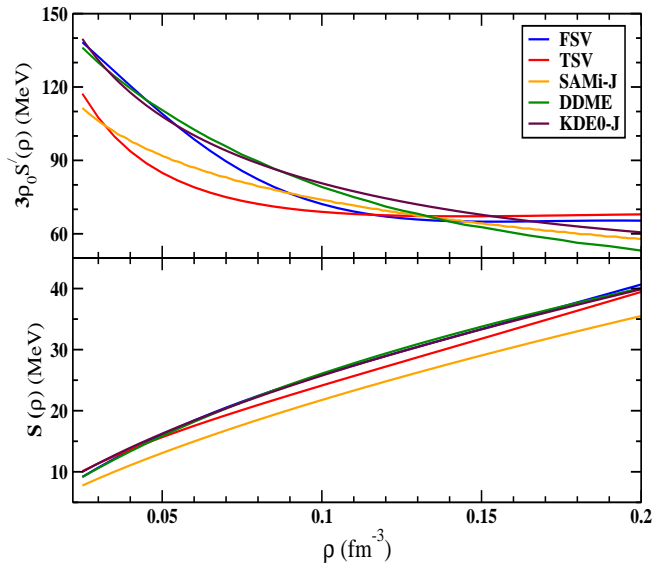}}
\caption{\label{fig5} (Color online) 
The nuclear symmetry energy $S$ (lower panel) and its density derivative
$S'$ multiplied by $3\rho_0$ (upper panel)  as a function of density
for the five different models associated with the slope parameter for
nuclear matter $L$ $\sim 65$ MeV.  Each of these models belongs to a
different family (see also Table \ref{tab1}).}
\end{figure}

In an attempt to understand the issues raised at the end of the previous
subsection, we make a detailed comparison between the results for the
five models belonging to different families but yielding almost the
same values for $L$. We first take a closer look in Fig. \ref{fig5}
into the values of the symmetry energy $S(\rho)$ (lower panel) and its
density derivative $3\rho_0S^\prime(\rho)$ (upper panel) as a function
of density for these models. The behavior of $S(\rho)$ as a function of
density seemingly appears to be similar for the five models. But the
values of $3\rho_0S^\prime(\rho)$ show significant differences in the
low density region ($\rho<0.10\ \text{fm}^{-3}$). Furthermore, one may
note that the TSV and SAMi-J models corresponding to $\Delta r_{\rm
np}(^{208}\text{Pb}) \sim 0.18$ fm and the KDE0-J model with $\Delta
r_{\rm np}(^{208}\text{Pb}) \sim 0.19$ fm display a relatively similar
behavior in the density dependence of $S^\prime(\rho)$.  The same
is true for the FSV and DDME models corresponding to $\Delta r_{\rm
np}(^{208}\text{Pb}) \sim 0.22$ fm.

To investigate whether such differences in the values of the density
derivative of the symmetry energy at lower densities have an influence
in the finite nuclei calculations, and motivated by Eq. (\ref{eq:Asym}),
we determine an effective value of the slope parameter $L_{\rm eff}$,
which might be more sensitive to the relative distributions of neutrons
with respect to protons in finite nuclei, as follows:
\begin{equation}
 L_{\rm eff} = \frac{3\rho_0\int \left[r^2\rho(r)I^2(r)\right]S^\prime(\rho(r))dr} {\int \left[r^2\rho(r)I^2(r)\right]dr}.
\label{leff}
\end{equation}
Here, $I(r)$ is the local asymmetry parameter defined as, $I(r)
\equiv (\rho_n(r)-\rho_p(r))/\rho(r)$. If one assumes $S(\rho)$ to
be linear in density, the $L_{\rm eff}$ parameter coincides with $L$
(see Eq. (\ref{s_rho})). However, we have seen in Fig. \ref{fig5}
that $S(\rho)$ can depart significantly from linearity at low
densities. Therefore, the $L_{\rm eff}$ parameter as defined in
Eq. (\ref{leff}) tries to take into account this effect. At very low
densities ($\rho < 0.01\ {\rm fm}^{-3}$) $S(\rho)$ deviates largely
from linearity. The integrals in the numerator and denominator of
Eq. (\ref{leff}) are thus evaluated by integrating from the center of the
nucleus, where the density $\rho(r)$ is of the order of $\rho_0$, up to
the point where the density of the nucleus falls to 0.01 fm$^{-3}$, which
corresponds to a radial coordinate $r$ of about 9 fm. It is worthwhile
to mention that we wanted to study the effect of $S^\prime(\rho)$
but not the quantity $L(\rho)\ (\equiv 3\rho S^\prime(\rho))$ on the
$\Delta r_{\rm np}$ of a heavy nucleus. That is why we kept $\rho_0$
outside the integral of the numerator in Eq. (\ref{leff}). The values of
$L_{\rm eff}$ along with various other properties evaluated for the five
models corresponding to $L\sim 65$ MeV are compared in Table \ref{tab1}.
\begin{table}[h]
\caption{\label{tab1} Comparison of the properties of infinite
nuclear matter (NM) and of the $^{208}$Pb and $^{132}$Sn nuclei for the
five different models that yield a value of $L$ around 65 MeV.}
 \begin{ruledtabular}
\begin{tabular}{ccccccc}
&	& SAMi-J	&	TSV	&	FSV	&	DDME & KDE0-J\\		
\hline
NM&$\rho_0$(fm$^{-3}$)&0.157&0.147	&0.149	&0.152& 0.162\\
&$L$(MeV)	& 63.2& 	67.0		&	64.8	& 	62.1& 65.7\\
&$J$(MeV)	&	30.00&	31.29		&33.16	& 	34.00& 35.00\\
$^{208}$Pb&$a_{\rm sym}(A)$(MeV)	& 20.35	&	22.20	&22.28& 	23.15& 24.18\\
&$\Delta r_{\rm np}$(fm)	& 	0.181	&	0.178&		0.223	& 	0.217& 0.188\\
&$\Delta r_{\rm np}^{\rm bulk}$(fm)& 0.109	&	0.086	&	0.119	& 	0.120& 0.108\\
&$L_{\rm eff}$(MeV)	& 81.2& 	82.7		&	95.7	& 	96.5& 90.8\\
$^{132}$Sn&$a_{\rm sym}(A)$(MeV)	& 19.24	&	21.27	&21.25& 	22.13& 23.06\\
&$\Delta r_{\rm np}$(fm)	& 0.245	&		0.239&		0.289	& 	0.279& 0.249\\
&$\Delta r_{\rm np}^{\rm bulk}$(fm)& 0.165	&	0.130	&	0.163	& 	0.165& 0.163\\
&$L_{\rm eff}$(MeV)	& 84.3& 	85.7		&  101.2	& 	98.0& 97.8\\
\end{tabular}
\end{ruledtabular}
\end{table}

It can be easily observed in Table \ref{tab1} that though the values
of $L$ for these models vary only by $\sim$\,5 MeV, the values of
$\Delta r_{\rm np}$ of heavy nuclei calculated from the same models
can differ by $\sim$\,0.05 fm, which is larger than the average spread
of the correlation between $\Delta r_{\rm np}$ and $L$. Interestingly,
when we look at the extracted $L_{\rm eff}$ parameter, the models from
SAMi-J and TSV families those predict $\Delta r_{\rm np}(^{208}\text{Pb})
\sim 0.18$ fm give similar $L_{\rm eff}\sim 82$ MeV, and the models from
FSV and DDME families those predict $\Delta r_{\rm np}(^{208}\text{Pb})
\sim 0.22$ fm give similar $L_{\rm eff}\sim 96$ MeV. The model from
the KDE0-J family with $\Delta r_{\rm np}(^{208}\text{Pb}) \sim 0.19$
fm predicts $L_{\rm eff}\sim 91$ MeV.  That is, the models with larger
$L_{\rm eff}$ give larger $\Delta r_{\rm np}$ and vice versa. In fact,
further inspection of Fig. \ref{fig4} reveals that two members of
the FSV and DDME families with $\Delta r_{\rm np}(^{208}\text{Pb})
\sim 0.18$ fm, same as the SAMi-J and TSV models in Table \ref{tab1},
predict departing $L$ values ($L=53.2$ MeV in the FSV model and $L=46.5$
MeV in the DDME model). It turns out that these FSV and DDME models
also explore similar values of $L_{\rm eff}$ (83.9 MeV in FSV and 86.6
MeV in DDME) as done by the models from the SAMi-J and TSV families
displayed in Table \ref{tab1} with $\Delta r_{\rm np} \sim 0.18$ fm.
In principle, one can also define $L_{\rm eff}$ without the $I^2(r)$
terms in Eq. (\ref{leff}). That is why, we repeated the calculations
of $L_{\rm eff}$ by taking $I^2(r)$ to be unity in Eq.  (\ref{leff})
and found similar trends as explained above. In Table \ref{tab1},
concerning the properties of uniform matter, it is also noticeable that
the models do not display the same value of the saturation density. For
the non-relativistic functionals belonging to the SAMi-J and KDE0-J
family this value is about 5--10$\%$ larger than the values explored by
the relativistic functionals. This fact has some impact on the extracted
values of $L_{\rm eff}$ for these models (see Eq. (\ref{leff})).

\begin{figure}[]
\includegraphics[height=3.5in,width=3.2in,angle=-90]{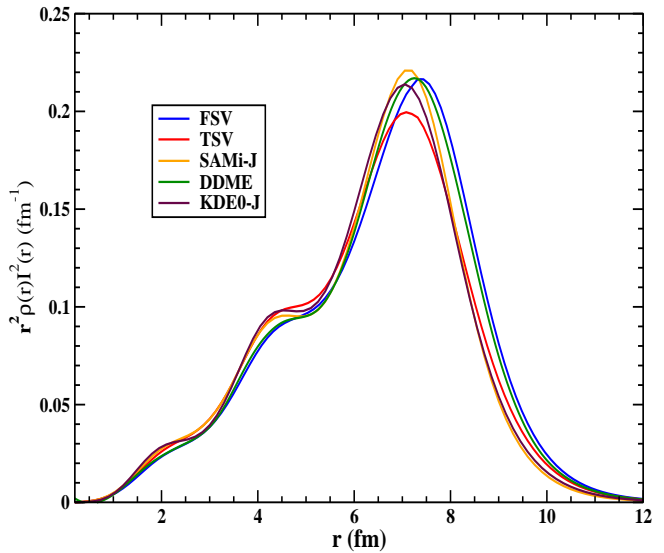}
\caption{\label{fig6} (Color online) The variation of $r^2\rho(r)I^2(r)$
as a function of the radial coordinate $r$ in $^{208}$Pb for the five
models that yield a symmetry energy slope parameter $L \sim 65$ MeV.}
\end{figure}

To have a better insight into the source of the differences between
the values of $L_{\rm eff}$ for the models with similar values of $L$
at $\rho_0$, we plot in Fig. \ref{fig6} the total density distribution
$\rho(r)$ of $^{208}$Pb multiplied by $r^2I^2(r)$ for the models with
$L\sim 65$ MeV. The values of $r^2\rho(r)I^2(r)$ for all the different
cases are close to each other up to $r \sim 6$ fm, in this region
$\rho(r) \geqslant 0.1 \text{ fm}^{-3}$. With further increase in $r$,
the differences in the values of $r^2\rho(r)I^2(r)$ gradually become
noticeable.  One can argue that different behaviors in the surface region
may be responsible for different values of $L_{\rm eff}$ and consequently
lead to different values of $\Delta r_{\rm np}$ in heavy nuclei like
$^{208}$Pb or $^{132}$Sn. The question still remains whether $L_{\rm eff}$
is more sensitive to the density dependence of $S^\prime(\rho)$ (upper
panel of Fig. \ref{fig5}) or to the density distributions of nucleons
inside the nucleus (Fig. \ref{fig6}). To unmask this, we calculated the
values of $L_{\rm eff}$ using $S^\prime(\rho)$ of a given model, but with
the density distributions of nucleons from the five models that have
$L\sim$ 65 MeV. We repeated this calculation for the different choices
of $S^\prime(\rho)$ of these five models. The values of $L_{\rm eff}$
so obtained did not show the trend as observed in Table \ref{tab1},
where $S^\prime(\rho)$ and the density distributions of nucleons used
correspond to the same model consistently.  Thus, the values of $L_{\rm
eff}$ are sensitive to both the density dependence of the symmetry
energy and the density distributions of nucleons inside the nucleus. To
this end, we would like to point out that the differences in the values
of $L_{\rm eff}$ for the models with similar $L$ parameter are mainly
due to the differences in the low density behavior of $S^\prime(\rho)$
and the distributions of nucleons in the surface region of the nucleus.

\section{Summary}

In this work, we revisit the correlations of the neutron-skin thickness
in finite nuclei with various symmetry energy parameters pertaining
to infinite nuclear matter. Particular attention is paid to the model
dependence in such correlations that can play a role in understanding
the density dependence of the nuclear symmetry energy. The finite
nuclei analyzed are $^{208}$Pb and $^{132}$Sn. The symmetry energy
parameters considered are $J - a_{\rm sym}(A)$, $a_{\rm sym}(A)/J$ and
$L$, where $J$ and $L$ are the symmetry energy and the symmetry energy
slope associated with infinite nuclear matter at the saturation density,
and $ a_{\rm sym}(A)$ corresponds to the symmetry energy parameter
in finite nuclei. Five different families of systematically varied
mean-field models corresponding to different energy density functionals
are employed to calculate the relevant quantities for the finite nuclei
and those for the infinite nuclear matter.  Consideration of recent
constraints on the symmetry energy parameters ($ a_{\rm sym}(A)$ and
$J$) and the present correlations suggest the values $\Delta r_{\rm np}=
0.197 \pm 0.047$ fm and $\Delta r_{\rm np}= 0.260 \pm 0.050\,{\rm fm}$
for the neutron skin thickness in the $^{208}$Pb and $^{132}$Sn nuclei,
respectively and $L=64\pm23$ MeV.

In general, the correlations of the neutron-skin thickness with the
different symmetry energy parameters are strong within the individual
families of the models. Once the results for all the different families
are combined, the correlation coefficients become smaller, indicating
a model dependence. The neutron skin in a nucleus entails two main
components related to the geometry of the nucleon density profiles. On
the one hand, there is a bulk contribution ($\Delta r_{\rm np}^{\rm
bulk}$) produced by the separation between the effective sharp surfaces
of the density distributions of neutrons and protons. On the other
hand, there is a surface contribution ($\Delta r_{\rm np}^{\rm surf}$)
caused by the different surface widths of the neutron and proton density
profiles. The correlations of the symmetry energy parameters with the
bulk part $\Delta r_{\rm np}^{\rm bulk}$ of the neutron-skin thickness
are less model dependent than with the total neutron-skin thickness
$\Delta r_{\rm np}$.  Exceptionally, the bulk part of the neutron-skin
thickness is correlated with $J - a_{\rm sym}(A)$ and $a_{\rm sym}(A)/J$
in an almost model independent manner. This fact is much compatible with
the predictions of the macroscopic droplet model.

We notice a model dependence in the correlations of the neutron-skin
thickness with the symmetry energy slope parameter $L$ when the results
of the various families of models are considered together. By model
dependence we mean that different models of different families with
the same value of the slope $L$ of the symmetry energy predict different
neutron skin thickness, or vice versa. For different models having similar
slope parameter $L \sim$ 65 MeV and belonging to the different families, a
spread in $\Delta r_{\rm np}$ of about 0.05 fm is observed, which is large
in view of the average spread of the correlation (Fig. (\ref{fig4})),
as well as in view of the DM estimate for the change of $\Delta r_{\rm
np}$ with $L$.

We have found two independent indications that the surface of the
nucleus plays a key role in introducing a model dependence, or in other
words, a systematic theoretical uncertainty, to the well-known linear
correlation between the neutron skin thickness and $L$ and to some other
correlations that can be used to extract the parameters characterizing
the density dependence of the symmetry energy. These indications are,
(i) the existence of stronger correlations of various symmetry energy
parameters with the bulk part of the neutron-skin thickness rather
than with the total neutron-skin thickness, and (ii) the differences
between the density distributions for the nucleons at the surface region
for the different models corresponding to similar values of the slope
parameter $L$.

To understand better the model dependence in the various correlations
considered, the results are compared for the models belonging to different
families, but yielding similar values of $L$. We have determined an
effective value of the symmetry energy slope parameter $L_{\rm eff}$
using the density distributions of nucleons and the density derivative
of the symmetry energy for these models. It is found that the values of
$\Delta r_{\rm np}$, which differ for the models with the same $L \sim$
65 MeV, are in harmony with the values of $L_{\rm eff}$. We conclude
that differences in the values of $L_{\rm eff}$ caused by differences
in the density distributions of nucleons in the surface region and the
derivative of the symmetry energy at subsaturation densities are the
plausible sources for the aforesaid model dependence.

\section*{Acknowledgments}

G.C. would like to thankfully acknowledge the nice hospitality
extended to him during his visit to SINP, when this work has
started. N.P. acknowledges support from FP7-PEOPLE-2011-COFUND program
NEWFELPRO, the Croatian Science Foundation under the project Structure
and Dynamics of Exotic Femtosystems (IP-2014-09-9159) and the QuantiXLie
Center of Excellence. M.C. and X.V.  acknowledge partial support from
Grant No.\ FIS2014-54672-P from the Spanish MINECO and FEDER, Grant No.\
2014SGR-401 from Generalitat de Catalunya, the Consolider-Ingenio 2010
Programme CPAN CSD2007-00042, and the project MDM-2014-0369 of ICCUB
(Unidad de Excelencia Mar\'{\i}a de Maeztu) from MINECO.


\end{document}